\documentclass[%
 reprint,
 amsmath,amssymb,
 aps,
floatfix,
]{revtex4-2}

\usepackage{graphicx}
\usepackage{dcolumn}
\usepackage{bm}
\usepackage[usenames,dvipsnames,svgnames,table]{xcolor}
\usepackage{array}
\usepackage{multirow}
\usepackage{soul}
\usepackage[T1]{fontenc}
\usepackage{selinput}
\usepackage{booktabs}

 \def\be{\begin{equation}}
 \def\ee{\end{equation}}
 \def\bea{\begin{eqnarray}}
 \def\eea{\end{eqnarray}}
 \def\bean{\begin{eqnarray*}}
 \def\eean{\end{eqnarray*}}
\begin{document}

\preprint{APS/123-QED}

\title{Deuteron Production in Ultra-Relativistic Heavy-Ion Collisions:\\A Comparison of the Coalescence and the Minimum Spanning Tree Procedure}

\author{Viktar Kireyeu$^{1,2}$}

\author{Jan Steinheimer$^{3}$}

\author{J\"org Aichelin$^{3,4}$}

\author{Marcus Bleicher$^{2,5,6,7}$}

\author{Elena Bratkovskaya$^{2,5,6}$}

\affiliation{$^{1}$ Joint Institute for Nuclear Research, Joliot-Curie 6, 141980 Dubna, Moscow region, Russia}
\affiliation{$^2$ Helmholtz Research Academy Hessen for FAIR (HFHF), GSI Helmholtz Center for Heavy Ion Physics, Campus Frankfurt, 60438 Frankfurt, Germany}
\affiliation{$^3$ Frankfurt Institute for Advanced Studies, Ruth Moufang Str. 1, 60438 Frankfurt, Germany}
\affiliation{$^4$ SUBATECH, Universit\'e de Nantes, IMT Atlantique, IN2P3/CNRS
4 rue Alfred Kastler, 44307 Nantes cedex 3, France}
\affiliation{$^5$ Institut f\"ur Theoretische Physik, Johann Wolfgang Goethe-Universit\"at,
Max-von-Laue-Str. 1, 60438 Frankfurt am Main, Germany}
\affiliation{$^6$ GSI Helmholtzzentrum f\"ur Schwerionenforschung GmbH,
  Planckstr. 1, 64291 Darmstadt, Germany} 
\affiliation{$^7$ John von Neumann-Institut f\"ur Computing, Forschungzentrum J\"ulich,
52425 J\"ulich, Germany}

\date{\today}

\begin{abstract}
The formation of deuterons in heavy-ion collisions  
at relativistic energies is investigated by employing two recently advanced 
models -- the Minimum Spanning Tree (MST) method and the coalescence model by 
embedding them in the  PHQMD and the UrQMD transport approaches. 
While the coalescence mechanism combines nucleons into deuterons at the kinetic freeze-out hypersurface, the MST identifies the clusters during the different stages of time evolution. 
We find that both clustering procedures give very similar results for the deuteron observables in the UrQMD as well as in the PHQMD environment. Moreover, 
the results agree well with the experimental data on deuteron production in Pb+Pb collisions at $\sqrt{s_{NN}} = 8.8$ GeV (selected for the comparison of the methods and models in this study). 
A detailed investigation shows that the coordinate space distribution of the produced deuterons differs from that of the free nucleons and other hadrons. Thus, deuterons are not destroyed by additional rescattering.

\end{abstract}

\maketitle

\section{Introduction}
The observation of light baryonic clusters in the central rapidity region in ultra-relativistic heavy-ion collisions is a topic of current interest with substantial research activity, both on the theoretical side and on the experimental side. A central physical question that emerged during that last years is how such weekly bound objects could be produced in and survive the hot and dense environment created in the central collision region. A central methodological question is how to identify/calculate such clusters in (dynamical) simulations of heavy-ion reactions.

Early on, models for the description of light nuclei production in nuclear collisions have been developed \cite{Schwarzschild:1963zz,Butler:1963pp,Kapusta:1980zz,Bond:1977fd}. These clustering models were later improved using the Wigner function and phase space coalescence approach  \cite{Gyulassy:1982pe,Aichelin:1987rh,Nagle:1994wj,Nagle:1996vp,Ko:2010zza,Botvina:2014lga,Zhu:2015voa,Botvina:2016wko,Sombun:2018yqh,Zhao:2021dka,Sun:2020uoj,Glassel:2021rod} to allow for the calculation of clusters from dynamical simulations. The consensus here was that nuclear clusters are formed in the late dilute (but still baryon rich) stage of the collision, through nuclear interactions. Later on, light nuclei where also detected in the more energetic collisions at high energy ion colliders. Prime examples are the observation of clusters by the STAR collaboration at RHIC (Relativistic Heavy-Ion Collider) in Brookhaven \cite{Adam:2019wnb,Arsene:2010px}, as well as by the ALICE collaboration at the LHC (Large Hadron Collider)  \cite{Adam:2015vda,Acharya:2017dmc} and by the NA49 Collaboration at the SPS (Super Proton Synchrotron) \cite{NA49:2010lhg,Anticic:2016ckv} at CERN, Geneva.

In fact, statistical model fits, based on the multiplicity of a multitude of hadrons, yield a freeze out temperature of the clusters of about $T=160$ MeV and a system which is dominated by mesons \cite{Andronic:2017pug}. According to the results of lattice gauge calculations this temperature is close to the pseudo critical temperature of the chiral crossover at vanishing chemical potential, which separates the phase where hadrons dominate from that where partons are dominant. In such an environment clusters, which have a binding energy of a couple of MeV per nucleon, would not be stable and cannot survive until the kinetic freeze-out which can be determined by the transverse momentum spectra of the hadrons produced in these collisions, which show an inverse slope parameter of more than 100 MeV. This apparent discrepancy has been dubbed the 'snowball in hell' puzzle.

Besides the dynamical formation and coalescence at kinetic freeze-out \cite{Schwarzschild:1963zz,Butler:1963pp,Kapusta:1980zz,Bond:1977fd,Nagle:1996vp,Scheibl:1998tk,Ko:2010zza,Botvina:2014lga,Botvina:2016wko,Sombun:2018yqh,Zhao:2021dka,Sun:2020uoj,Glassel:2021rod}, further descriptions for light nucleus production have been discussed in the literature. Because the thermal fit to hadron multiplicities almost perfectly describes also the nuclei yields even at the highest beam energies \cite{Andronic:2017pug} an instant chemical freeze-out at the pseudo critical temperature has been proposed. However, such a scenario seems not compatible with the finding that the kinetic freeze out of nuclei coincides with that of nucleons and pions which takes place at a much lower temperature, around 100 MeV. Recent works that also describe the multiplicity and flow of nuclei are either based on partial chemical equilibrium models, where e.g. the deuteron yield is maintained throughout the kinetic rescattering phase \cite{Bebie:1991ij,Vovchenko:2019aoz,Xu:2018jff,Neidig:2021bal}, or explicit scattering processes which can form and destroy light nuclei in a dynamical hadronic cascade model \cite{Oliinychenko:2018ugs,Oliinychenko:2020znl}. In general one can conclude that all of these approaches, including the dynamical formation and coalescence, seem to provide a reasonable description of experimental data. For most of them the formation of the observable nuclei occurs either at or after kinetic freeze-out, i.e. after the 'snowball has left hell'. 

Two methods that are widely employed to identify clusters in dynamical simulations are the Minimum Spanning Tree (MST) method (usually applied at lower collision energies) and coalescence (usually applied at higher collision energies). The goal of this study is to investigate deuteron production by comparing the results of these two different procedures, coalescence and MST, for cluster identification, if they are applied to the microscopic PHQMD and UrQMD transport approaches. For this we compare the deuteron multiplicity as well as differential observables (as the rapidity distributions and the $p_T$ spectra) obtained in UrQMD and PHQMD when applying the same cluster finding algorithm to both approaches. 
In practise we extended UrQMD and PHQMD simulation codes in a way that the MST method (previously used in PHQMD) and the coalescence method (previously used in UrQMD) are realised in a model independent cluster recognition psMST library \cite{Kireyeu:2021igi} which can then be applied on equal footing to both transport approaches.

Furthermore, we investigate whether the coalescence procedure leads  
also to the same conclusion - as found in \cite{Glassel:2021rod} -  that clusters  can survive the expansion phase since they move behind the main stream of free nucleons and hadrons. 

For the present comparative study we focus on Pb+Pb collisions at $\sqrt{s_{NN}} = 8.8$ GeV, where experimental data, measured by the NA49 collaboration \cite{Anticic:2016ckv}, are available.

The paper is organized as follows:
In Section II we briefly evoke the basic ideas of the cluster recognition procedures -  coalescence and MST - implemented in the PHQMD and UrQMD models. 
In Section III we present the results for deuteron production in heavy-ion collisions.
In Section IV we study the time evolution of deuteron production.
Finally, in Section V we summarize our findings.

\section{Cluster production in dynamical models of heavy-ion collisions}

In general, one might assume that the calculation of nuclear clusters in relativistic nuclear collisions is rather straight forward. Based on the full n-body (time dependent) quantum Wigner distribution of the nucleons, the Hamiltonian of the system should allow for the binding of nucleons into nuclear clusters using the (spin and iso-spin) dependent potential interactions. These clusters should be stable asymptotic states that can be identified at $t\rightarrow\infty$. Such approaches are possible at very low collision energies and are realized e.g. in fermionic/anti-symmetrized Molecular Dynamics models \cite{Feldmeier:2016zut,Myo:2017gjc},  but are out of reach for beam energies well above 1 AGeV, due to the relativistic kinematics.

At relativistic energies, one usually employs for the deuteron production a phase space approximation to the quantum Wigner density of the deuteron. Such approaches (like the PHQMD and the UrQMD) simulate the n-body phase space average over spin states and include only 2-body (density dependent) interactions. The QMD based models can provide a satisfactory description of the underlying n-body phase space distributions of nucleons to be used as a basis for cluster production. While one expects that all simulation models provide the same results for the nucleon distribution, due to differences in detail, like the parametrization of unmeasured cross sections, and due to the complexity of the dynamics involved, the results differ in detail but they tend to give a qualitatively similar though quantitatively differing results. In order to take into account and estimate such uncertainties we employ two different models in two different setups for the cluster production, namely the PHQMD model and the UrQMD model, either in so-called ''cascade mode'' or including the interactions of nucleons via hadronic potentials (potential mode).

\subsection{The Parton-Hadron-Quantum-Molecular Dynamics (PHQMD)} The PHQMD approach \cite{Aichelin:2019tnk,Glassel:2021rod} as well as the Ultra-relativistic-Quantum-Molecular Dynamics (UrQMD) \cite{Bass:1998ca,Bleicher:1999xi}
describe ultra-relativistic heavy-ion collision using Quantum Molecular Dynamics. In these approaches the nucleons are presented as Wigner densities of the wave functions. The n-body Wigner density is the direct product of the one body Wigner densities (at the energy discussed here anti-symmetrization can be neglected). The time evolution of the n-body Wigner density is given by a variational principle. Because the nucleons interact by mutual density dependent two-body interactions, energy and momentum are strictly conserved. 
The n-body QMD approaches \cite{Aichelin:1991xy,Hartnack:2015vzc} allow then to study the correlations (e.g. a deuteron as a 2-body correlation). 

This distinguishes the QMD from the BUU type approaches like SMASH, AMPT or GiBUU, which propagate only the 1-body phase space density $f_1(\vec r,\vec p)$ in a mean field by using the test particle method. 
There, the 2-body phase space density  is factorized into the product of two 1-body densities  neglecting the correlation term:
$f_2(\vec r_1,\vec p_1,\vec r_2,\vec p_2,)=f_1(\vec r_1,\vec p_1)\cdot f_1(\vec r_2,\vec p_2)$.  
This factorization is a major short coming, because it rules out the calculation of genuine 2-body correlations, like deuterons.  
 
A similar problem is present for the Parton-Hadron-String Dynamics (PHSD)  approach for strongly interaction systems  \cite{Cassing:2008sv,Cassing:2009vt,Cassing:2008nn} based on  the Kadanoff-Baym equations  \cite{KadanoffBaym} in first-order gradient expansion \cite{Cassing:2008nn}. It propagates density matrices (in phase-space representation) which contain information not only on the occupation probability (in terms of the phase-space distribution functions as in BUU), but also on the properties of hadronic and partonic degrees-of-freedom via their spectral functions (cf. Ref. \cite{Cassing:2021fkc}). However, the nucleon-nucleon potential in the PHSD approach is realized also on the mean-field level and therefore prevents the direct study of deuteron production.

To overcome this limitation, the PHQMD approach was developed to include full n-body dynamics for the expansion stage of the reaction. PHQMD is based on the PHSD approach (in version 4.0 which includes the description of the partonic phase \cite{Cassing:2008sv,Cassing:2009vt,Bratkovskaya:2011wp,Linnyk:2015rco}). 
The  QGP phase in PHQMD is described in terms of strongly interacting quasiparticles based on the DQPM (Dynamical Quasi-Particle Model). As introduced in Ref.~\cite{Aichelin:2019tnk}, the cluster identification can then be realized in PHQMD  by the  Minimum Spanning Tree (MST) procedure \cite{Aichelin:1991xy} (also used in this study) or by the Simulated Annealing Clusterization Algorithm (SACA) (see \cite{Puri:1996qv,Puri:1998te}), as it was previously done in QMD \cite{Puri:1996qv,Gossiaux:1997hp,PhysRevC.100.034904} calculations.

\subsection{The Ultra-relativistic-Quantum-Molecular Dynamics (UrQMD)} 

UrQMD \cite{Bass:1998ca,Bleicher:1999xi} includes as degrees of freedom the full set of established hadrons and their resonances, summing up to more than 50 baryon species and 40 meson species. The parametrization of the cross sections among these hadrons are similar to, but differ in details from that in PHQMD. UrQMD allows to include, similar to PHQMD,  soft and hard (density dependent) two body interactions, which are mostly relevant at low beam energies, while at higher energies the model is usually used in the so called 'cascade mode' which means that between collisions or after decays hadrons are propagated on straight line trajectories.

For the present study we use the UrQMD and PHQMD models in two modes: the cascade mode without potential and the potential mode including nucleon-nucleon potentials (cf. Ref. \cite{Hillmann:2018nmd} for UrQMD). In both models the potential is a density dependent Skyrme like potential with a hard equation-of-state.

However, there is a difference in the practical realization of the potential in both approaches. This difference is related to the fact that a covariant formulation of 
the relativistic QMD equations requires the reduction from a 8-dimensional space (covariant $x^\mu=(t,\vec r)$ and $p^\mu=(E,\vec p)$) to 6+1 dimensions (i.e. ($\vec r, \vec p, t$)) by employing energy and time constraints \cite{Marty:2012vs}. At the moment the full numerical realization of this method for realistic heavy-ion calculations is not feasible. Therefore, in order to extend the QMD approach to relativistic energies, in the PHQMD a modified single-particle Wigner density of the nucleons has been introduced 
(cf.  \cite{Aichelin:2019tnk}) to account for the Lorentz contraction of the nucleus in the beam $z$-direction in coordinate and momentum space by the inclusion of $\gamma_{cm} =1/\sqrt{1-v_{cm}^2}$, where $v_{cm}$ is the velocity of the
bombarding nucleon in the initial $NN$ center-of-mass system.   
Accordingly, the interaction density is modified, what leads to the modification
of the 2-body forces between nucleons. 

In the UrQMD realization, there is no additional $\gamma_{cm}$ modification included, i.e. single particle densities are treated as spherical in coordinate space. This leads to different strengths of the potentials and the results depend on this different implementation. Let us stress, that if we use the same single particle densities in PHQMD and UrQMD, the results of the both involved models are very similar.

As a side remark, also a QGP phase can be included into the UrQMD dynamics by switching to a hydrodynamical hybrid description \cite{Steinheimer:2007iy,Petersen:2008dd} for the most dense stages of the reaction. For this study, however, we will use only the hadron-string dynamics of the UrQMD discarding the QGP formation, i.e. without a transition to the hydro mode. 

In the past the coalescence procedure has been applied to study deuteron production in the UrQMD approach for various energies, see e.g. \cite{Sombun:2018yqh,Hillmann:2019wlt,Gaebel:2020wid,Kittiratpattana:2020daw,Hillmann:2021zgj}.

\section{Cluster formation in the PHQMD and UrQMD}

In this section we briefly recall the basic ideas of the Minimal Spanning Tree (MST) and coalescence procedures for the cluster recognition as (previously) employed in the PHQMD and UrQMD microscopic transport approaches.

\subsection{Cluster recognition by the minimum spanning tree (MST)}
In the QMD  approach (here specifically the PHQMD realization) nucleons interact by potentials and by collisions. 
The potential between nucleons is attractive around nuclear ground state densities and therefore at the end of the heavy-ion reaction nucleons tend to stay together and to form clusters. To identify clusters a Minimum Spanning Tree (MST) procedure is applied, which will be described in the following: In the implementation of the MST algorithm used in this study, only the coordinate space information is used to identify clusters. A nucleon is considered as part of a cluster if its spatial distance to any other nucleon is less than $r_0 = 4$~fm in the local rest frame of the cluster. The distance is calculated by a Lorentz transformation from the computational frame to the local rest frame and the cut-off distance is chosen according to the  range of the potential in PHQMD. Nucleons which are more distant than the cut-off distance are assumed to be not bound by the attractive nuclear interaction of this specific cluster. 
The main advantage of MST is to allow to identify clusters at any time during the evolution of the system. The clusters, created by the potential interaction during the time evolution, are therefore formed dynamically.  

We have checked that additional cuts in momentum space change the result only marginally  (cf. \cite{Kireyeu:2021igi}) because clusters with a relative momentum larger than the corresponding binding energy do not stay together but separate in the expanding system.
Being a semi-classical simulation of the bound cluster states, the clusters are not fully stable as a function of time and the cluster multiplicity has to be determined at a given time. It has been checked that a determination of the cluster states at different (earlier or later) times change only the multiplicity of the clusters but neither the form of the rapidity distribution nor that of the transverse momentum distribution. 

From a previous analysis of the cluster production in PHQMD, the conclusion has been advanced that clusters are closer to the center of the reaction as compared to free nucleons and hadrons \cite{Glassel:2021rod}. The fact that nucleons in a cluster and free hadrons are not at the same place in coordinate space may therefore explain why collisions between clusters and hadrons are not frequent and therefore do not destroy the produced clusters during the further evolution of the system.

\subsection{Phase space coalescence approach}

The phase space coalescence approach is a statistical description of cluster production, based on proximity in momentum and coordinate space. It can also be employed in n-body approaches when the underlying kinetic description of the hadron dynamics does not include the dynamical formation of nuclei, for example through nuclear potential interactions. This is a main reason for the popularity of the coalescence model in the simulation of high energy nucleus-nucleus collisions, because such models do often not include a nuclear potential. A coalescence procedure has been recently applied in Refs.
\cite{Botvina:2014lga,Sombun:2018yqh} to the UrQMD approach
and in Refs. \cite{Ko:2010zza,Zhu:2015voa,Chen:2003ava,Sun:2020uoj,Sun:2021dlz,Lin:2003ah} to the AMPT model
(a multi-phase transport model).

In the present coalescence implementation for deuteron production, the relative positions and momenta for all proton-neutron pairs in their center-of-mass (CM) frame is calculated after the individual last scattering of the two nucleons. The following procedure is employed to generate deuterons: I) The time of last scattering of each of the both nucleons in the p-n pair is calculated from the simulation. The last scattering of the pair is defined by the nucleon of the pair which scatters at the later time.
II) The p-n center-of-mass frame is determined and the nucleon, which had its last collision earlier is propagated to the freeze-out time of the other nucleon.
III) The relative momentum $\Delta P$ and distance $\Delta R$ between the proton and the neutron in its CM frame is calculated. If $\Delta P<0.285$ GeV and $\Delta R< 3.575$ fm, a deuteron may be formed.
IV) If the above condition is met, then the probability that a deuteron is formed is given by the spin-isospin combinatorial factor $P_d=3/8$.
V) If a deuteron is formed its momentum is given by the sum of the $p-n$ momenta in the CM frame, boosted back into the computational frame.

Note, that in this approach the two coalescence factors $\Delta P$ and $\Delta R$ are considered as free parameters and are fixed once and then used to make predictions at other systems and beam energies. Other coalescence approaches try to constrain the free parameters by assuming the width of the deuteron wave function in phase space \cite{Ko:2010zza,Zhu:2015voa,Chen:2003ava}. A comparison between both coalescence approaches showed that they yield within 30\% similar results as long as the product $\Delta P \Delta R$ is fixed \cite{Nagle:1996vp}, which is within the systematic error obtained from the variation of the deuteron wave function (Hulthen wave function vs. harmonic oscillator wave function) in the Wigner function approach. 

For further details on the procedure to identify clusters -  MST and coalescence -
we refer the reader to the original publications of the PHQMD \cite{Aichelin:2019tnk} and UrQMD \cite{Botvina:2014lga,Botvina:2016wko,Sombun:2018yqh} groups. 
Furthermore, the detailed description of differences between the PHQMD (and PHSD) and the UrQMD transport approaches can be found in a recent review \cite{Bleicher:2022kcu}.

\section{Bulk observables for deuteron production}

In this section, we present the results of the study for the "bulk" observables, such as rapidity distributions and transverse momentum spectra.
In order to explore the influence of the nucleon-nucleon potential on the deuteron production we apply two modes of the PHQMD and the UrQMD transport approaches: in the first mode we switch-off the nucleon-nucleon potential (cascade mode), while in the second mode the standard nucleon-nucleon potential is active (potential mode).

As mentioned in Section II, in the PHQMD approach in the regions of high local energy density (above a critical energy density of $\varepsilon_C \simeq 0.5$\  GeV/fm$^3$)
a transition from hadronic to partonic degrees-of-freedom occurs, i.e. the system is in the QGP phase. Here, the nucleon-nucleon potential acts on baryons  only during the hadronic phase, i.e. during the primary $NN$ collisions, after the hadronization 
and on baryons from the hadronic 'corona' in the peripheral reaction region,
which are not part of the QGP.
We use UrQMD in the hadron-string mode (i.e., without a QGP phase realized via hydrodynamics). Therefore, in UrQMD the nucleon-nucleon potential is also active when the system is strongly compressed and dense. 

We mention that nucleons, which do not scatter at all, are not included in the analysis of the deuterons. They are only relevant for the fragmentation region, where the dissociation of the spectator should be better treated with a multi-fragmentation approach and not via coalescence or MST. For the results at midrapidity, studied in this paper at $\sqrt {s_{NN}}=8.8$ GeV, the fragmentation region can be well separated and is not relevant for this investigation.

\begin{figure}[!t]
    \centering
    \resizebox{0.45\textwidth}{!}{
        \includegraphics{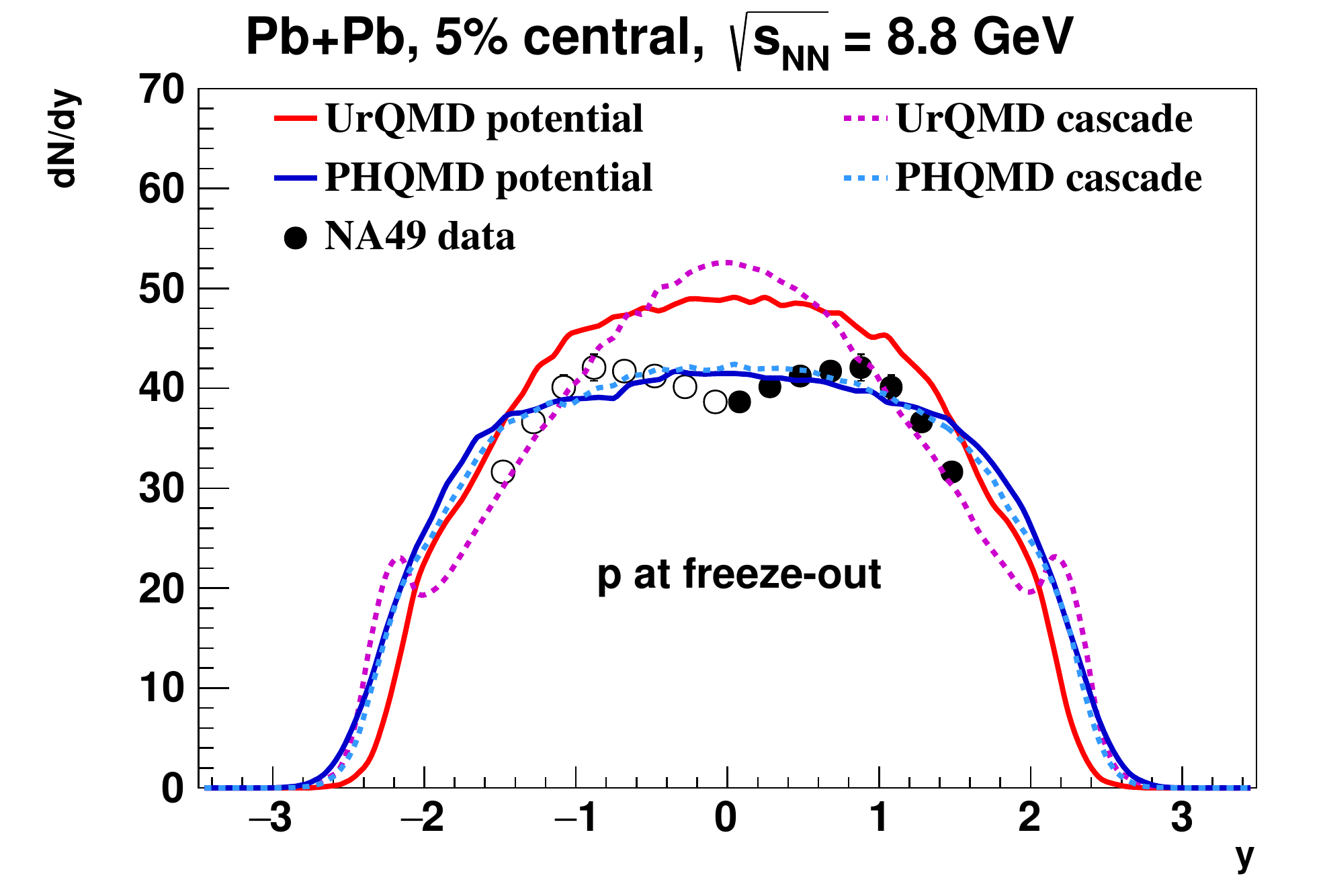}
    }
    \caption{Rapidity distribution of protons at the "freeze-out" time before the coalescence. The solid red line corresponds to the UrQMD with potential, the magenta dashed line represents UrQMD results in the cascade mode, the blue solid line shows PHQMD predictions with potential, while the dashed cyan line is for PHQMD in cascade mode. The experimental data (black circles) are from NA49 Collaboration \cite{NA49:2010lhg}.}
    \label{pn:rapidity}
\end{figure}

\begin{figure}[!ht]
    \centering
    \resizebox{0.45\textwidth}{!}{
        \includegraphics{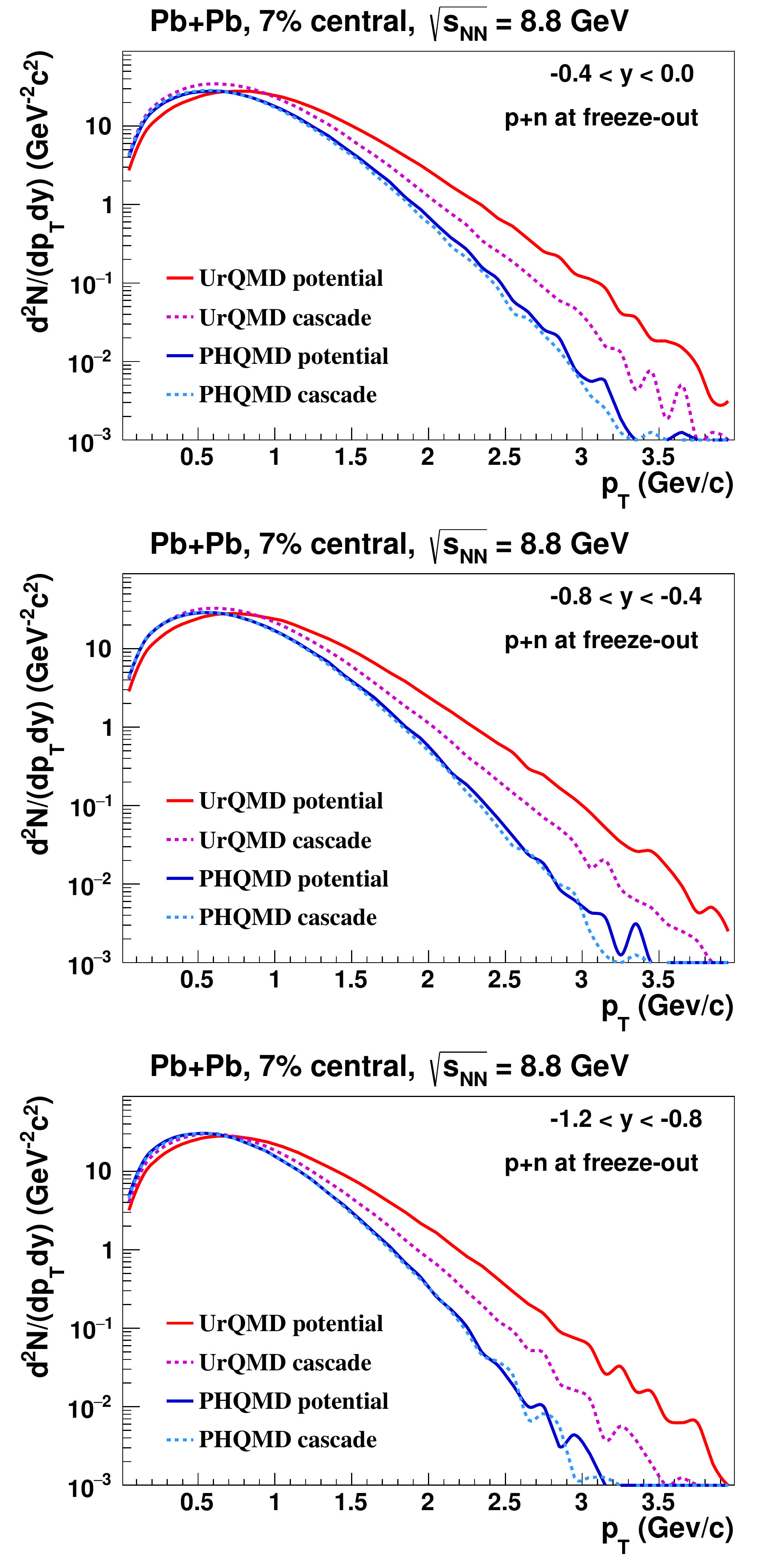}
    }
    \caption{Transverse momentum distribution of nucleons ($p+n$) at the "freeze-out" time before the coalescence in the rapidity intervals of $-0.4 < y < 0.0$ (top), $-0.8 < y < -0.4$ (middle) and $-1.2 < y < -0.8$ (bottom) . UrQMD results are shown in red color, PHQMD in blue color. The cascade mode calculations are denoted by dotted lines and potential mode simulations by full lines.}
    \label{p:pt}
\end{figure}

\subsection{Proton rapidity and $p_T$ distributions}

Since the dynamics of the nucleons is essential for the deuteron production, we start with the investigation of the proton rapidity distributions.
In Fig. \ref{pn:rapidity} we compare the rapidity distributions of all protons, which had at least one collision, with the data from the NA49 collaboration \cite{NA49:2010lhg} for central Pb+Pb collisions ($b\le 3$ fm) at $\sqrt {s_{NN}}=8.8$ GeV calculated at the  "freeze-out" time before the coalescence is applied.

The solid red line corresponds to the UrQMD with potential, the magenta dashed line represents UrQMD results in the cascade mode, the blue solid line shows PHQMD predictions with potential, while the dashed cyan line is for PHQMD in cascade mode. The experimental data (black circles) are from NA49 Collaboration \cite{NA49:2010lhg}. 

The baryon number is strictly conserved in both approaches but 
there is a differences in the final baryon chemistry: UrQMD has  about 15\%  more protons then PHQMD. This is compensated by a larger number of strange baryons in PHQMD.
For PHQMD the difference between cascade mode and potential mode is negligible whereas UrQMD in cascade mode leads to a higher central rapidity density. The reason for the increased rapidity density in UrQMD in case of the cascade mode is due to the fact that the nucleon-nucleon potential is repulsive at high densities which leads in consequence to a weaker stopping, stronger transverse flow and a broader rapidity distribution in case of the potential calculation.

Next we turn to the transverse momentum distribution of the nucleons (protons+neutrons) at freeze-out, but before coalescence as shown in Fig. \ref{p:pt}. From top to bottom we show the transverse momentum distributions for different rapidity windows defined by the deuteron measurements of the NA49 collaboration discussed below (top: $-0.4<y<0$, middle: $-0.8<y<-0.4$, bottom: $-1.2<y<-0.8$ ) for the two different models (UrQMD in red color, PHQMD in blue color) each in cascade mode (dotted lines) and potential mode (full lines). We observe here as well that PHQMD in the cascade mode yields the same spectrum as in the potential mode. The UrQMD spectra are generally stiffer as compared to the PHQMD spectrum. In the UrQMD in the potential mode the spectrum is stiffer than in the cascade mode  due to the release of the nucleon-nucleon potential energy which is built up during the high density phase of the heavy-ion reaction.

It is clear from the discussion above that both models, in different modes, provide visibly different phase space distributions already for the protons and neutrons. Quantitatively, the differences for the bulk of the matter, i.e. at central rapidities and transverse momenta below 1.5 GeV are on the order of 15-30\% . We see this as a systematic uncertainty of the underlying distributions which consecutively will form deuterons. It is therefore expected that, even if the formation mechanism (coalescence or MST) is identical in the models, the final deuteron spectra for the different models may also deviate.

\begin{figure}[!t]
    \centering
    \resizebox{0.45\textwidth}{!}
    {\includegraphics{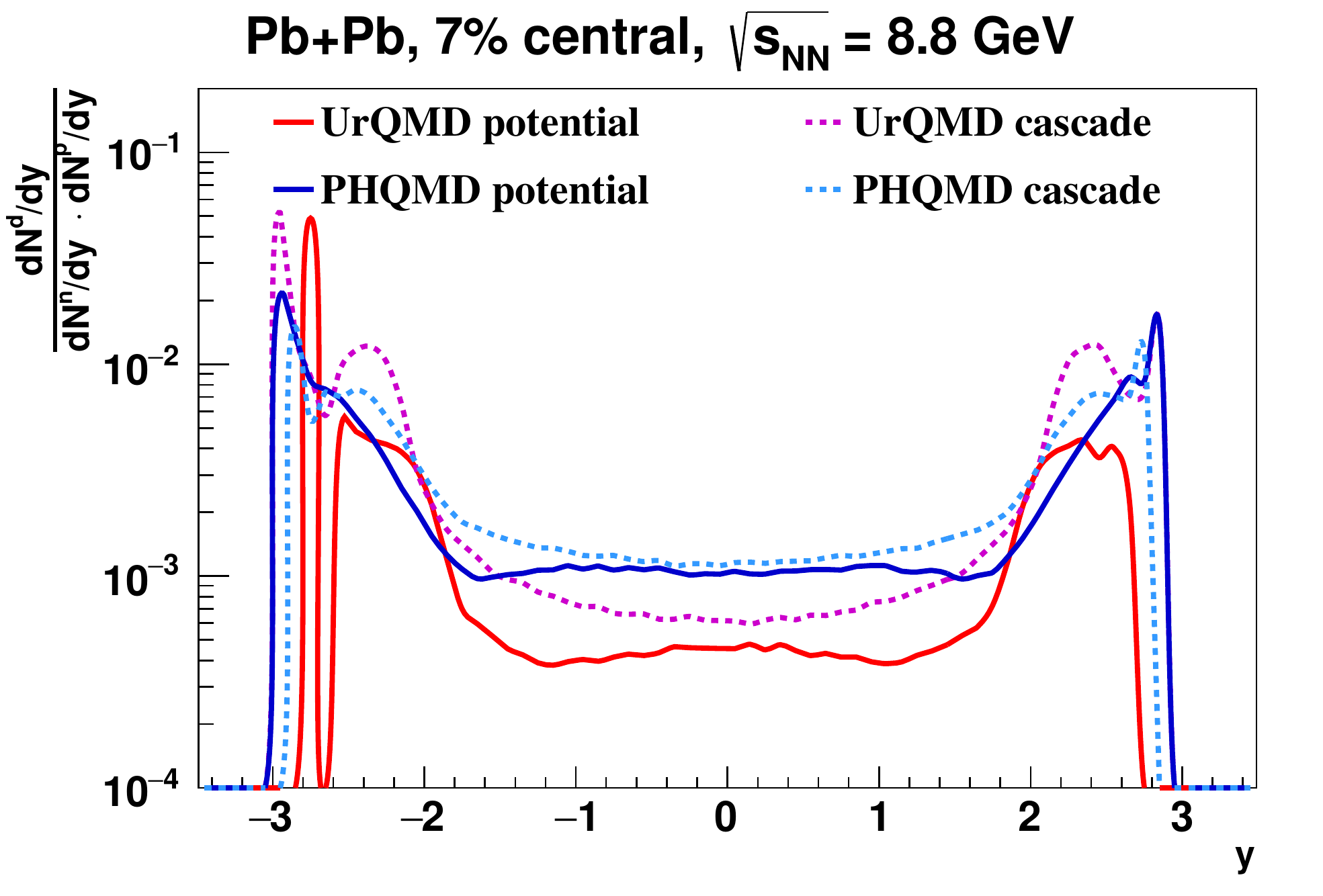}}
    \caption{The deuteron to proton times neutron ratio $
    (dN^d/dy)/{(dN^n/dy \cdot dN^p/dy) }$ as a function of rapidity. UrQMD results are shown in red color, PHQMD in blue color. The cascade mode calculation are denoted by dotted lines and potential mode simulations by full lines. Here, all calculations use the coalescence approach for deuteron production.
    }
    \label{b2_rapidity}
\end{figure}

\subsection{Deuteron rapidity and $p_T$ distributions}

The rapidity distribution of deuterons, normalized to the product of the neutron and proton rapidity distribution is shown in Fig. \ref{b2_rapidity} for PHQMD (blue) and UrQMD (red) and for the two different options for the potential. For this comparison both models use the coalescence approach to obtain the clusters.
We see that UrQMD produces less deuterons, per proton-neutron pair, at midrapidity than PHQMD. This hints to different freeze out conditions in the two models, i.e. differences in the relative spatial and momentum coordinates of protons and neutrons at their last scattering. More precise, UrQMD seems to have a lower phase space density of protons and neutrons at kinetic freeze out which then translates into a lower coalescence probability.
In both approaches the shape of the distribution is similar for cascade and for the potential set up respectively. 

\begin{figure*}[!ht]
    \centering
    \resizebox{\textwidth}{!}{
        \includegraphics{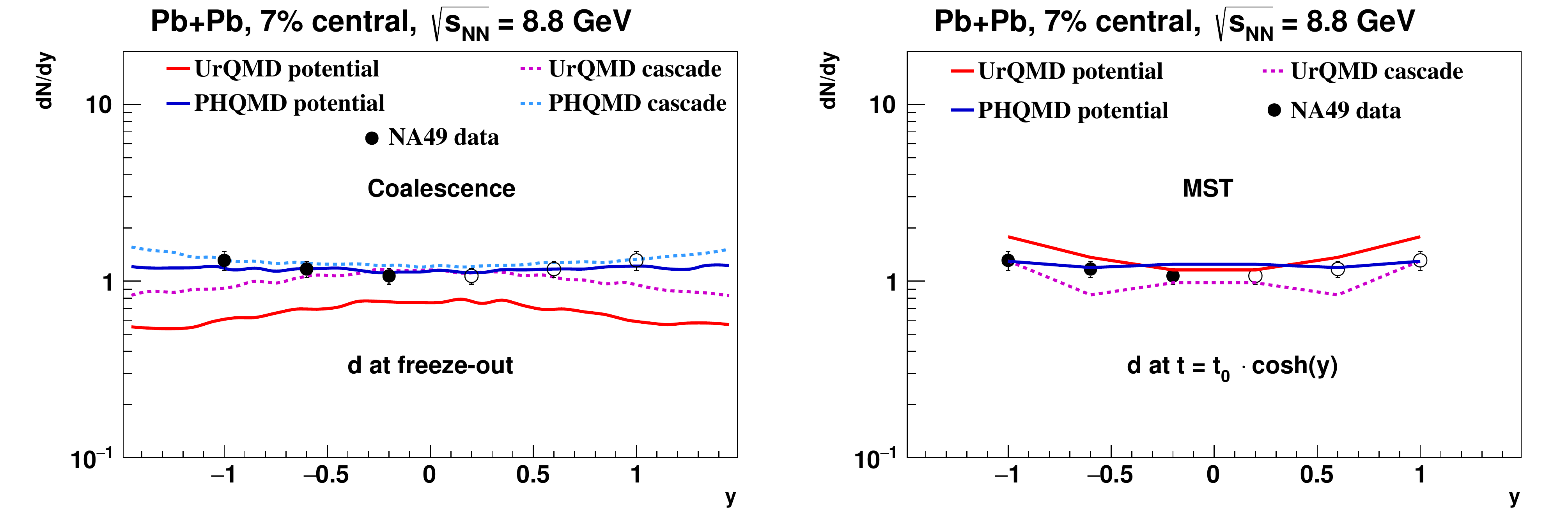}
    }
\caption{Rapidity distribution of deuterons in central Pb+Pb reaction at $\sqrt{s_{NN}}=8.8$ GeV. UrQMD results are shown in red color, PHQMD in blue color. The cascade mode calculation are denoted by dotted lines and potential mode simulations by full lines. Left: Distribution of deuterons found by the coalescence algorithm at the freeze-out time. Right: Distribution of deuterons found by the MST algorithm for the time $t = t_{0} \cosh(y)$. The experimental data (black circles) are from the NA49 Collaboration \cite{Anticic:2016ckv}. 
}
    \label{deut:rapidity}
\end{figure*}

The final deuteron rapidity distributions for central Pb+Pb collisions at $\sqrt {s_{NN}}=8.8$ GeV are displayed in Fig.  \ref{deut:rapidity} and compared with the experimental data from the NA49 Collaboration \cite{Anticic:2016ckv}. UrQMD results are shown in red color, PHQMD in blue color. The cascade mode calculations are denoted by dotted lines and potential mode simulations by full lines, the experimental data is mirrored and shown as circles. On the left hand side we show the rapidity distribution of deuterons obtained by applying the coalescence procedure, on the right hand side the deuterons are calculated by applying the MST procedure at fixed eigentime $t_0$. Following the procedure of Ref. \cite{Glassel:2021rod} for the MST we extract the yield at the physical time $t$ with $t=t_0 \cosh(y)$, where $t_0$ is the time measured in the computational frame, the nucleus-nucleus center of mass frame. $t_0$ is different for the different modes
($t_{0} = 100$ fm/c for UrQMD potential mode $t_{0} = 60$ fm/c for UrQMD cascade mode and $t_{0} = 50$ fm/c for PHQMD (both modes)). We observe that both methods, coalescence and MST, to identify deuterons give very similar results, for the multiplicity as well as for the form of the rapidity distribution. 

\begin{figure*}[!ht]
    \centering
    \resizebox{\textwidth}{!}{
        \includegraphics{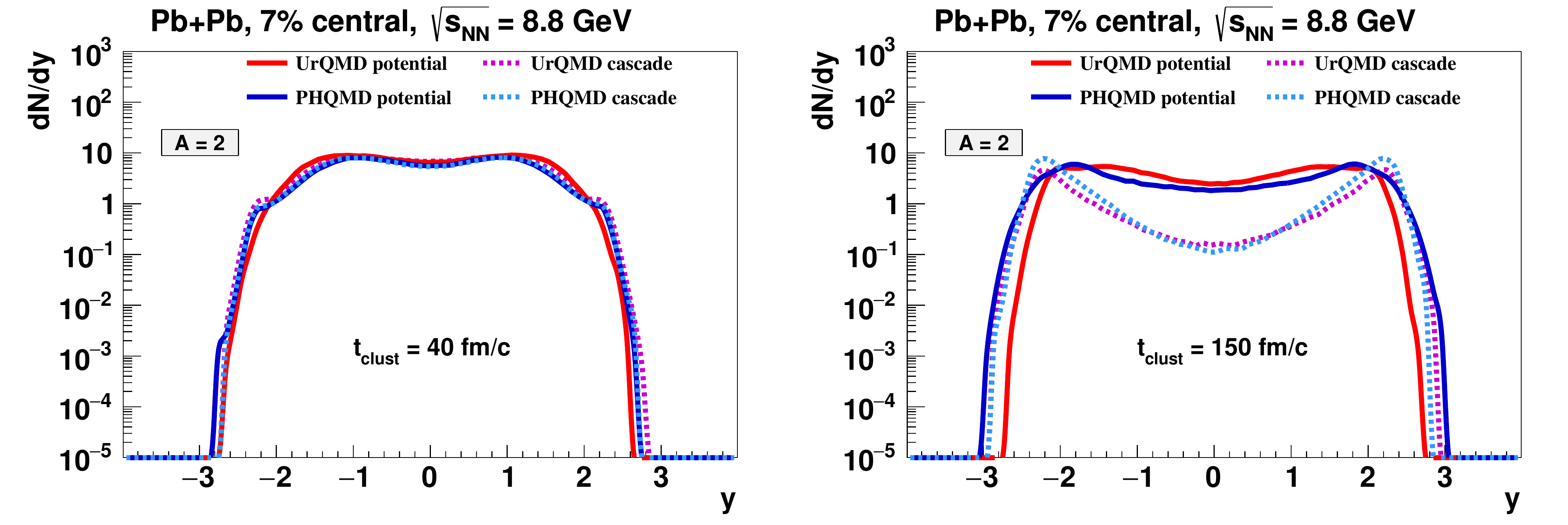}
    }
    \caption{Rapidity distributions for deuterons (clusters with the mass number $A=2$) 
    identified within the MST procedure in central Pb+Pb reaction at $\sqrt{s_{NN}}=8.8$ GeV. UrQMD results are shown in red color, PHQMD in blue color. The cascade mode calculations are denoted by dotted lines and potential mode simulations by full lines. Left: $t_{clust} = 40$ fm/c, right: $t_{clust} = 150$ fm/c.}
    \label{mst:a2}
\end{figure*}

To illustrate the dependence of the deuteron distribution on the nucleon-nucleon potential  Fig. \ref{mst:a2} shows the result of the MST analysis for two fixed times:
$t_{clust} = 40$ fm/c (left figure) and $t_{clust} = 150$ fm/c (right figure). The left figure demonstrates that the 2-particle correlations even at such rather late times are similar in the calculations with and without potential interaction, because the yields and the shape of the deuteron distribution are nearly identical between all four simulations. However, when going to later times, one observes that only simulations including the nucleon-nucleon potential are able to keep the clusters bound (which means they keep the 2-particle nucleon-nucleon correlation), while for the simulations without potential the deuteron yield decreases at central rapidities. This can be understood straightforward, because without the attractive nucleon-nucleon potential interaction, the proton and neutron increase their spatial distance with time and can not form (or be identified as) deuterons any longer.

\begin{figure*}[!ht]
    \centering
    \resizebox{\textwidth}{!}{
        \includegraphics{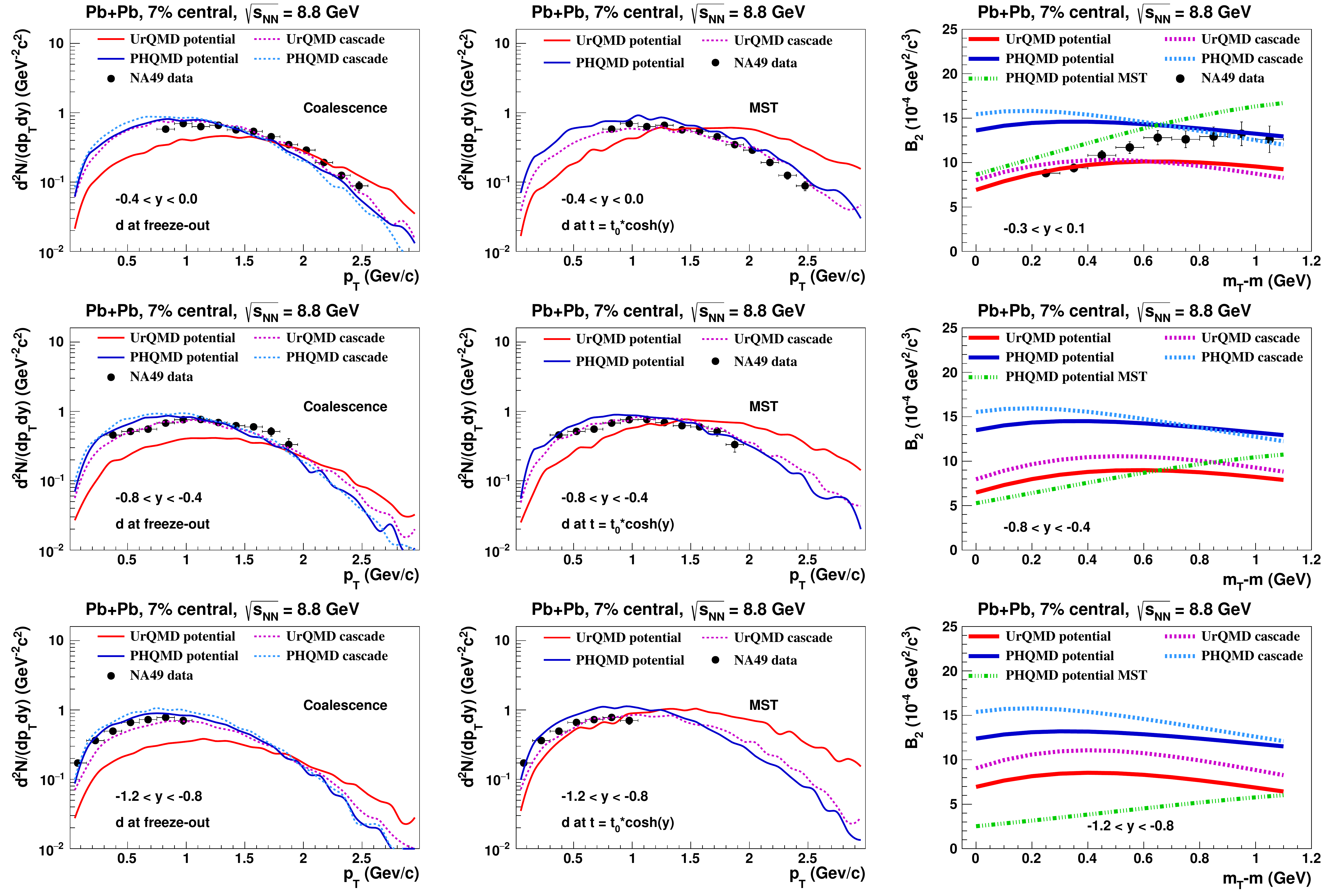}
    }
    \caption{Transverse momentum distribution of deuterons in central Pb+Pb reaction at $\sqrt{s_{NN}}=8.8$ GeV. UrQMD results are shown in red color, PHQMD in blue color. The cascade mode calculations are denoted by dotted lines and potential mode simulations by full lines. The  experimental data (black circles) are from the NA49 Collaboration \cite{Anticic:2016ckv}. The left column shows the calculations using coalescence, the middle column the MST results and the right column shows the coalescence factor $B_2$ versus transverse mass $m_{T}-m_0$ in the rapidity interval $-0.3 < y < 0.1$ using coalescence and an additional line (green dashed dotted) showing the PHQMD with potential calculation using the MST. From top to bottom we show the rapidity bins: $-0.4 < y < 0.0$ (top row), $-0.8 < y < -0.4$ (middle row) and $-1.2 < y < -0.8$ (bottom row).     }
    \label{deut:pt}
\end{figure*}

In Fig. \ref{deut:pt} we address the transverse momentum distribution of deuterons in central Pb+Pb reaction at $\sqrt{s_{NN}}=8.8$ GeV. UrQMD results are shown in red color, PHQMD in blue color. The calculations in cascade mode are denoted by dotted lines and those in potential mode by full lines. The  experimental data (black circles) are from the NA49 Collaboration \cite{Anticic:2016ckv}. The left columns shows the calculations using coalescence, the middle column the MST results and the right column shows the coalescence factor $B_2$
\be
B_2 = \frac{E\frac{{\rm d}^3N_d}{{\rm d}P_d^3}}{\Big(E\frac{{\rm d}^3N_{neutrons}}{{\rm d}p_n^3}\Big) \Big(E\frac{{\rm d}^3N_{protons}}{{\rm d}p_p^3}\Big)}.
\ee 
versus transverse mass $m_{T}-m_0$ in the rapidity interval $-0.3 < y < 0.1$ (as provided by the experiment) using coalescence. The green dash-dotted lines show for comparison the PHQMD results with potential interaction for the MST scenario calculated at freeze-out time at the same clusterization time as on the right panel of the Fig. \ref{deut:rapidity}. From top to bottom we show the rapidity bins: $-0.4 < y < 0.0$ (top row), $-0.8 < y < -0.4$ (middle row) and $-1.2 < y < -0.8$ (bottom row).  

For all rapidity bins we see a quite good agreement between the theoretical results and the experimental data, independent of the method to identify deuterons. Only for UrQMD with potential the spectra are stiffer due to the repulsion of the nucleon-nucleon potential at high density which is related to the non-relativistic Gaussian form of interaction density as mentioned in Section II. 

We can conclude from this section that the deuteron multiplicity as well as their rapidity and transverse momentum distribution is rather independent of the way in which the deuteron yield is obtained from the underlying nucleon distribution. The coalescence and MST procedures give essentially not only the same result but the predictions agree also well with the experimental data. In view of the very different underlying methodology of both procedures this is a remarkable result. 

\section{Space and time distribution of deuteron production}

As presented in Section III, the
deuterons, identified either by the coalescence model or by MST are in good agreement with the experimental data. Therefore it is useful to take advantage of the transport approaches and to study in details the production process. 
To be close to the experimental situation, we concentrate here on the midrapidity region $|y| < 1$.

\begin{figure*}[!ht]
    \centering
    \resizebox{\textwidth}{!}{
        \includegraphics{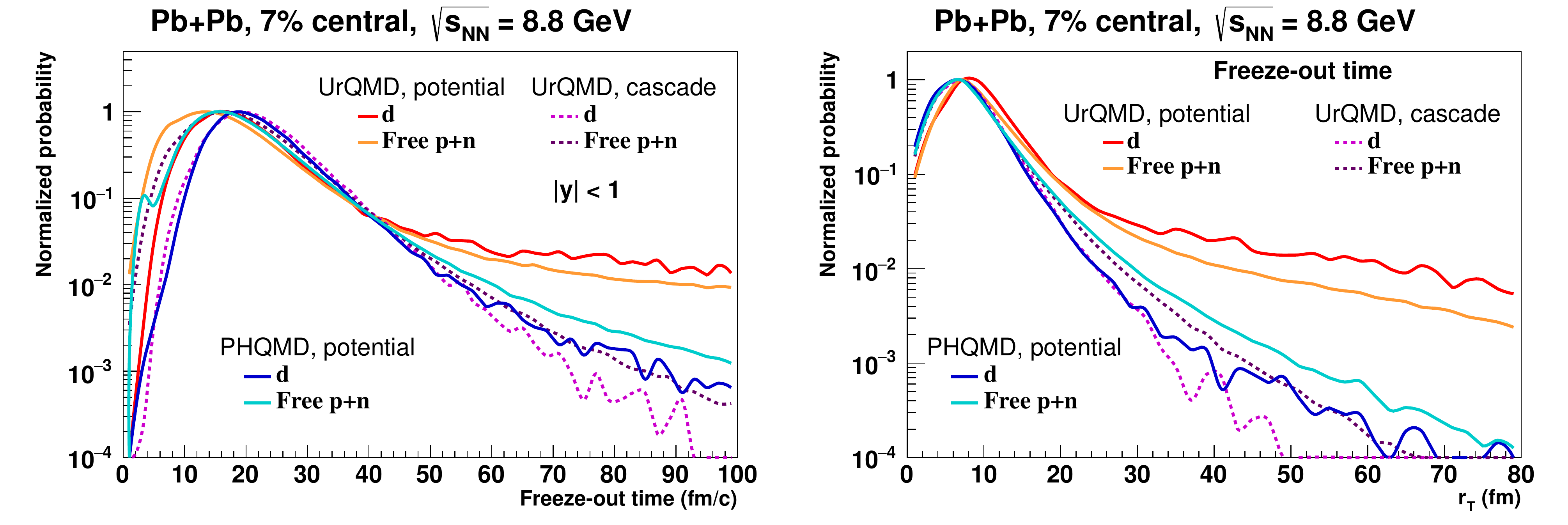}
    }
    \caption{Freeze-out time distribution (left) and transverse distance distribution at freeze-out (right) of free (i.e. unbound) nucleons ($p+n$) and deuterons (using coalescence).
    The solid red line shows the deuterons from UrQMD (potential mode), 
    the solid dark red line the free $p+n$ from UrQMD (potential mode), 
    the dashed green line represents deuterons from UrQMD (cascade mode), 
    the dashed dark green line the free $p+n$ from UrQMD (cascade mode), 
    the blue solid line shows the deuterons from PHQMD (potential mode), 
    the dark blue solid line the free $p+n$ from PHQMD (potential mode).}   
    \label{prod_time}
\end{figure*}

\begin{figure*}[!ht]
    \centering
    \resizebox{\textwidth}{!}{
        \includegraphics{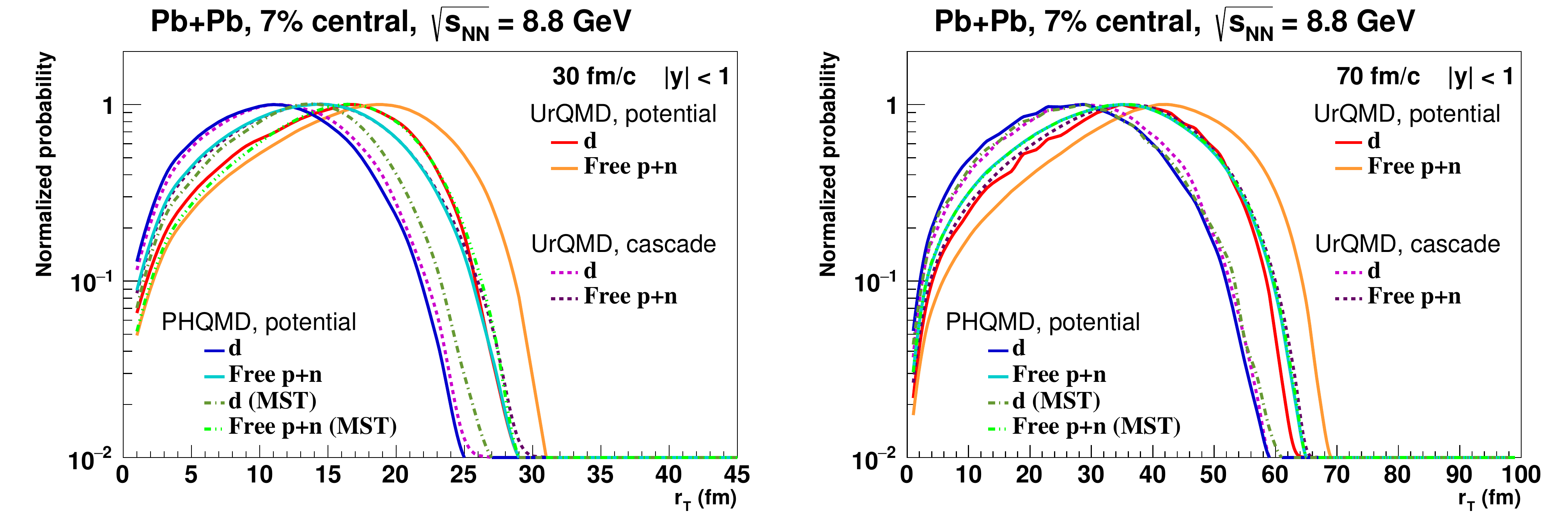}
    }
    \caption{Transverse distance of unbound nucleons ($p+n$) and deuterons at $30$ fm/c (left) and at $70$ fm/c (right). The color scheme is the same as on the Fig. \ref{prod_time}. Additionally, the dot dashed blue line shows PHQMD deuterons found by the MST algorithm.}
    \label{tr_dist_times}
\end{figure*}

To get an overview of the space time evolution of the heavy-ion reaction we start out with the distribution of the freeze-out time and transverse distance of nucleons and deuterons (using the coalescence model) as displayed in Fig. \ref{prod_time}. The solid red line shows the deuterons from UrQMD (potential mode), the solid dark red line -- the free $p+n$ from UrQMD (potential mode), the dashed green line represents the deuterons from UrQMD (cascade mode), the dashed dark green line the free $p+n$ from the UrQMD (cascade mode), the blue solid line shows the deuterons from  PHQMD (potential mode) and the dark blue solid line the free $p+n$ from the PHQMD (potential mode). We define the freeze-out time of each individual nucleon as the time at which this nucleon had its last (elastic or inelastic) collision. The freeze-out (production) time deuterons in the coalescence model, is defined by the freeze-out time of the coalescing nucleons and taken to be the later time of the two nucleons.  The distribution of the freeze-out times (Fig. \ref{prod_time}, left) are generally peaked around $10-20$ fm/c, however, we observe a systematic shift by approximately $5$ fm between the freeze-out time of free nucleons and that of nucleons which are bound in deuterons, which freeze-out later. At $40$ fm/c the freeze-out probability has decreased to 10\% of its peak value and practically all collisions have ceased. The curves are very similar for PHQMD and UrQMD in cascade mode. A peculiarity observed in the potential mode of UrQMD is that the bound nucleons continue to have very soft collisions apparently extending the freeze-out time. These collisions among nucleons bound in  clusters are suppressed in PHQMD by a minimal energy threshold. When turning to the transverse distribution of the nucleons and deuterons at freeze-out we observe that the majority (defined again by 10\% of the peak probability) nucleons and deuterons freeze-out within $15$ fm from the center of the reaction. Here we also observe the long tail for the UrQMD potential calculations which is due to the soft scattering discussed above.

Let us now explore the transverse spacial distribution of nucleons and deuterons in more detail. In Fig.  \ref{tr_dist_times} we show on the left hand side the normalized distribution at a fixed center-of-mass time of $30$ fm/c and on the right hand side at a fixed center-of-mass time of $70$ fm/c. One should note, that this time cut is different from the one used in Fig. \ref{prod_time} (right) where the location of the nucleons and deuterons at the freeze-out time was shown. The color and line style coding is the same as in the previous Fig.  \ref{prod_time}. Again, the deuterons are identified by the coalescence procedure. We show additionally for PHQMD the results for the MST procedure which gives nearly identical results as the coalescence and do not show the MST results for the other models to avoid to overcrowd the figure.

One clearly observes that the nucleon and deuteron positions follow the expansion flow with the maxima of the distributions shifting by approximately $20$ fm/c over the duration from $30$ fm/c to $70$ fm/c. An interesting observation is however that the deuterons remain at smaller radial distances than the nucleons (also the average $r_T$ of deuterons is in all transport approaches smaller than that of the free nucleons). We quantify this difference between nucleons and deuterons using the location of the peak value $P_{\rm max}$ and summarize the results in Tab.  \ref{table:rt:30} and \ref{table:rt:70} for fixed times of $30$ fm/c and $70$ fm/c. One should remember that the $B_2$ distribution, Fig. \ref{deut:pt}, is almost flat. The deuterons have therefore a similar velocity as the nucleons, but their distance to the reaction center, Fig. \ref{tr_dist_times}, is smaller than that of the free nucleons. Consequently, the deuterons are produced at a smaller distance to the reaction center than the free protons or probably at a later time. This is a hint for a production at the end of the hadronization of the QGP or from nucleons which are closer to the center of the reaction than the average.

\begin{table}[!ht]
\centering
\begin{tabular}{m{12em} c c}
\toprule
\multirow{2}{12em}{Model mode} & \multicolumn{2}{c}{$r_T$ (fm) at $P_{\rm max}$ } \\
  & d & p \\
\midrule
UrQMD potential & 17 & 19 \\
UrQMD cascade   & 11 & 15 \\
PHQMD potential & 11 & 15 \\
PHQMD potential (MST) & 13 & 17 \\
\bottomrule
\end{tabular}
    \caption{Transverse distance $r_T$ of nucleons and deuterons at $P_{\rm max}$  at $30$ fm/c. The table corresponds to the lines in Fig. \ref{tr_dist_times}, left.}
    \label{table:rt:30}
\end{table}

\begin{table}[!ht]
\centering
\begin{tabular}{m{12em} c c}
\toprule
\multirow{2}{12em}{Model mode} & \multicolumn{2}{c}{$r_T$ (fm) at $P_{\rm max}$ } \\
  & d & p \\
\midrule
UrQMD potential & 35 & 41 \\
UrQMD cascade   & 29 & 35 \\
PHQMD potential & 27 & 35\\
PHQMD potential (MST) & 29 & 35 \\
\bottomrule
\end{tabular}
    \caption{Transverse distance $r_T$ of nucleons and deuterons at $P_{\rm max}$ at $70$ fm/c. The table corresponds to the lines in Fig. \ref{tr_dist_times}, right.}
    \label{table:rt:70}
\end{table}

\section{Conclusion}
We have presented and compared the coalescence and Minimum Spanning Tree (MST) procedures to calculate deuteron production in the PHQMD and UrQMD models, both with and without nuclear potential interactions. The coalescence procedure assumes that deuterons are created when the last collision of the nucleons of the pair takes place and if at this time the relative distance of the two nucleons in coordinate and momentum space is lower than the coalescence parameters. In the MST procedure one follows the nucleons and determines at a much later (ideally asymptotically large) time whether the spacial distance between the proton-neutron pair is smaller than $r_0=4$ fm. If this is the case and no other baryon is closer than 4 fm to one of the nucleons of the cluster this pair is considered as a deuteron. Based on this set-up, we have calculated the multiplicities, rapidity spectra and transverse momentum spectra of nucleons and deuterons in central Pb+Pb at $\sqrt{s_{NN}}=8.8$ GeV for all model/clustering combinations.

Our main findings are the following:
\begin{itemize}
\item When applied to one of the transport codes the coalescence and the MST procedures  provide very similar multiplicities and very similar rapidity and $p_T$ distributions.
\item The deuteron rapidity distributions and $p_T$ spectra obtained from the PHQMD code are very similar to those obtain from the UrQMD code, i.e. the results are model independent.
\item The deuteron rapidity distribution and $p_T$ spectra 
from central Pb+Pb at $\sqrt{s_{NN}}=8.8$ GeV agree quite well with the experimental data of the NA49 Collaboration.
\item The coalescence as well as the MST procedure show that the deuterons remain in transverse direction closer to the center of the heavy-ion collision than free nucleons. The results of the transport approaches are therefore different from the statistical model assumptions that assume a homogeneous distribution of nucleons and deuterons.
\item In this geometry, the deuterons do not pass the expanding baryonic fireball, but are spatially separated which might explain why they are not destroyed by collisions with the hot fireball hadrons.
\end{itemize}

\begin{acknowledgements}
The authors acknowledge inspiring discussions with Ch. Blume,  G. Coci, S. Gl\"a{\ss}el, C.M. Ko, V. Kolesnikov, V. Voronyuk, and P. Hillmann.

This study is part of a project that has received funding from the European Union’s Horizon 2020 research and innovation program under grant agreement STRONG – 2020 - No 824093. Furthermore, we acknowledge support by the Deutsche Forschungsgemeinschaft 
(DFG, German Research Foundation): grant BR~4000/7-1, by the Russian Science Foundation grant 19-42-04101 and by the GSI-IN2P3 agreement under contract number 13-70. JS acknowledges support from the BMBF through the ErUM-data-pilot project as well as the Samson AG.
Further support was provided by the PPP Program of the DAAD.
VK acknowledges support by the JINR grant 22-102-06.
Computational resources were provided by the Center for Scientific Computing (CSC) of the Goethe University, by the "Green Cube" at GSI, Darmstadt and
the NICA LHEP offline computing cluster at JINR.
\end{acknowledgements}

\bibliography{main}

\end{document}